\documentclass[preprint,aps]{revtex4}

\usepackage{graphicx}% Include figure files
\usepackage{dcolumn}% Align table columns on decimal point
\usepackage{bm}% bold math
\begin{document}
%------------------------------------------------------------------------------
\title{Reply to Comment on ``Roughness of Interfacial Crack Fronts:
Stress-Weighted Percolaton in the Damage Zone"}
\author{Jean Schmittbuhl}
\affiliation{Departement de G{\'e}ologie, ENS,
24, rue Lhomond, F--75231 Paris C{\'e}dex 05, France}
\author{Alex Hansen}
\affiliation{Department of Physics, NTNU, N--7491 Trondheim, Norway}
\author{G.\ George Batrouni}
\affiliation{INLN, Univ.\ Nice-Sophia Antipolis, 
1361 Route des Lucioles, F--06560 Valbonne, France} 
\date{\today}
%--------------------------------------------------------------------
\maketitle

\begin{figure}
\includegraphics[width=3.5cm,angle=0]{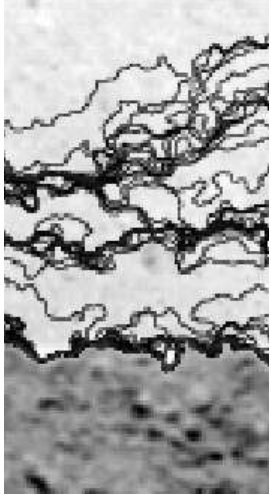}
\caption{Closeup of experimental fracture fronts from \cite{ms01}.} 
\label{fig1}
\end{figure}

Alava and Zapperi (AZ) \cite{az03} question whether the fracture fronts
we observe in \cite{shb03} are self affine.  We use 
Family-Vicsek scaling to determine the two scaling exponents $\alpha$ and
$z$.  AZ claim that this is not enough to determine whether the front is
self affine and they go on to point to the presence of overhangs as their
evidence of fractality rather than self affinity.  In Fig.\ \ref{fig1}, we 
show details of {\it experimental\/} fracture fronts from Fig.\ 1 of
\cite{ms01}.  There are significant overhangs, but these experimental fronts 
are self affine.  In fact, as long as one allows a damage cloud to develop, 
overhangs are unavoidable.  Family-Vicsek scaling implies non-isotropic 
scaling. A consequence is that the front width $w$ scales with the width 
of the system $L_x$.  Non-isotropic scaling is the defining property of 
self-affine surfaces.  

%--------------------------------------------------------------------
%--------------------------------------------------------------------
  
%--------------------------------------------------------------------
\begin{figure}
\includegraphics[width=6.5cm,angle=0]{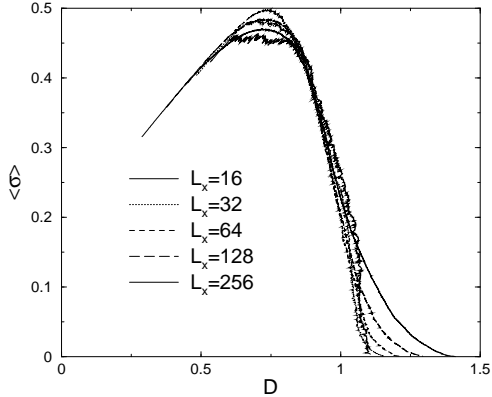}
\caption{Average stress $\langle \sigma\rangle$ as a function of
imposed displacement $D$ from virgin system to complete failure.
The rescaling of $\overline{G}_{i,j}$, Eq.\ (\ref{resc}) and of the threshold
distribution ensures that the curves collapse for different system lengths 
$L_x$ and fixed $L_y=128$.}
\label{fig2}
\end{figure}
%--------------------------------------------------------------------

AZ claim that the source of the $L_x$-dependency
of the damage length scale $l_y$, Eq.\ (9) in \cite{shb03}, comes from
the rescaling of the model's elastic constants when changing $L_x$.   
However, this 
rescaling is necessary to ensure that the elastic {\it properties\/} remain
unchanged when the system size is changed under uniform loading conditions.
It is caused by the non-local nature of the problem introduced by the
Green function $\overline{G}_{i,j}$, Eq.\ (4) in \cite{shb03}.  
Under uniform loading condition, Eq.\ (5) in \cite{shb03} reads 
$u_i=u = \sum_{j} \overline{G}_{i,j} f_j = \sum_{j} 
\overline{G}_{i,j} b^2 \sigma$.  Here $b$ is the lattice constant.
If $u$, local deformation, and $\sigma$, local
stress, are to be independent of size, $L_x$ and $L_y$, we must have that
\begin{equation}
\label{resc}
\sum_j \overline{G}_{i,j}={\rm constant}\;.
\end{equation}
Since $\overline{G}_{i,j}=\overline{G}_{i-j}$, we may for estetic reasons
write $\sum_j \overline{G}_{i,j}={\rm constant}=\sum_{i,j}
\overline{G}_{i,j}/(L_x \times L_y)$, as there is no dependency on the index 
$i$ in Eq.\ (\ref{resc}).  This was the way we chose to present it in
\cite{shb03}. We point out again that both indices in the double sum 
$\sum_{i,j}$ run over $L_x\times L_y$ sites.
We demonstrate the correctness of the rescaling in Fig.\
\ref{fig2}, where we show the collapse of the loading curves obtained for 
different system sizes after using Eq.\ (\ref{resc}).
Only when the elastic properties have been rescaled as just described,
one may proceed to use finite size scaling as done in \cite{shb03}.

We end this Reply by pointing out that Ramanathan and Fisher \cite{rf98}
measured numerically $\nu=1.52\pm0.02$ using a very different model,
which by construction cannot produce fractal fracture lines.  
This is in complete agreement with our model, which gave $\nu=1.54$. 
The fracture roughness exponents, however, are very different.

% --------------------------------------------------------------------

% --------------------------------------------------------------------
\end{document}